\documentclass[twocolumn,aps,showpacs]{revtex4}
\usepackage{amsmath,amsfonts}
\usepackage{amssymb}
\usepackage{theorem}
\usepackage{epsfig}
\newcommand{\Si}[1]{\ensuremath{#1_{sing}}}
\newcommand{\RE}[1]{\ensuremath{#1_{reg}}}
{\theorembodyfont{\rmfamily} \newtheorem{lemma}{Lemma}[section]}
\begin{document}
\title{\bf On stability of renormalized classical electrodynamics}
\author{Jerzy Kijowski}
\affiliation{ Center for Theoretical Physics,
Polish Academy of Sciences,  Aleja Lotnik\'ow 32/46, 02-668 Warsaw, Poland}
\author{Marcin Ko\'{s}cielecki}
\email{kosciej@fuw.edu.pl}
\affiliation{ Department of Mathematical Methods
in Physics, Warsaw University, ul. Ho\.{z}a 74, 00-682 Warsaw, Poland}
\date{\today}
\begin{abstract}
It is shown that the total energy of the static ``field +
particle'' system, defined in the framework of classical,
renormalized electrodynamics of particles and fields, depends in
an unstable way upon the field boundary data. It is argued that
this phenomenon may be also an origin of the unstable dynamical
behaviour of the system (i.e.~existence of ``runaway solutions'').
It is proved that a suitable polarization mechanism of the
particle restores the stability, at least on the level of statics.
Whether or not it restores also the full, dynamical stability of
the theory is still an open question.
\end{abstract}
\maketitle
\section{Introduction}

Classical electrodynamics in its present form is unable to
describe interaction between charged particles, intermediated by
electromagnetic field. Indeed, typical well posed problems of the
theory are of the contradictory nature: either we solve partial
differential equations for the field, with particle trajectories
providing sources (given {\em a priori} !), or we solve ordinary
differential equations for the trajectories of test particles,
with fields providing forces (given {\em a priori} !). Combining
these two procedures into a single theory leads to a
contradiction: Lorentz force due to self-interaction is infinite
in case of a point particle.

There were many attempts to overcome these difficulties. One of
them consists in using the Lorentz--Dirac equation (see
\cite{Dirac},\cite{Haag},\cite{Rohr}). Here, an effective force by
which the retarded solution computed for a given particle
trajectory acts on that particle is postulated (the remaining
field is finite and acts by the usual Lorentz force).
Unfortunately, this approach leads to the so called runaway
solutions which are unphysical.

Various remedies have been proposed to cure such disease,
 most of them just based on a fine tuning of boundary conditions.
Unfortunately, such a tuning excludes physically interesting
problems (i.e.~circular motion) and the question arises if one can
construct a theory which does not contain unphysical solutions at
all. The authors believe that to achieve the above goal we should
first gain a deeper understanding of foundations of the runaway
behaviour.

As a starting point of our analysis, we use an approach proposed
by one of us in papers \cite{EMP} and \cite{GKZ}. It consists in
defining an ``already renormalized'' four-momentum of the physical
system ``particle(s) + fields''. Equations of motion are then
derived as a consequence of the conservation law imposed on this
object. We deeply believe that such an approach is a a correct
realization of the Einstein's programme of ``deriving equations of
motion from field equations'' and that a similar procedure should
be applied to formulate the two-body-problem in General Relativity
Theory.

We show in the present paper, that the physical instability is
inherently contained in the renormalization method used. More
precisely: in the simplest renormalization scheme the amount of
energy contained ``in the interior of the particle'' decreases
when the external field surrounding the particle increases. This
contradicts the stability of the model. As a remedy for such
 drawback we propose the polarizability of the particle. Numerical
analysis of such an improved model shows validity of this
proposal.

In this paper we analyze the renormalized energy of the total
``particle + field'' system on the level of statics only, but the
energetic instability discovered this way is obviously a reason
for the runaway behaviour of the dynamical system as well. Indeed,
the price which must be paid for acceleration becomes negative.
This observation is fundamental, in our opinion, to understand the
physical reasons for the runaway behaviour of the theory and in
search for a remedy for this phenomenon.

The paper is organized as follows. In Section
\ref{renormalization} the renormalization procedure proposed by
one of us in \cite{EMP} (see also \cite{GKZ}) is presented. Then a
monopole particle inside a fixed volume $V$ is considered: we
compute renormalized energy of the system  and vary it with
respect to particle's position. Next, we assume that the particle
assumes position corresponding to minimal value of the energy. In
this way we obtain total energy of the system as a function of the
field boundary data, imposed on $\partial V$. Finally, we analyze
stability of the system under small changes of these data. Here,
both the Dirichlet-type and the Neumann-type boundary problems are
considered.

The above general results are then applied to a case of a monopole
particle closed in spherical box. We prove that such system {\em
is not} stable. Then we consider a polarizable particle. Here, the
external field may generate a non-vanishing dipole momentum, which
changes completely the energy balance. It turns out that for a
Heaviside-like relation between the field and the dipole momentum
it generates, the system is stable. This suggests a possible way
to improve in the future our renormalization method and to avoid
(maybe) also dynamical instabilities, manifesting themselves in
the runaway behaviour.



\section{The renormalized four-momentum
vector}\label{renormalization}

Full description of the renormalized electrodynamics was proposed
in \cite{EMP} or \cite{GKZ}. In the present Section we review
briefly heuristic ideas that stand behind definition of the
renormalized four-momentum of the dynamical ``particle + field''
system.

As a starting point of our considerations take an
extended-particle model. This means that we consider a fully
relativistic, gauge-invariant, interacting ``matter +
electromagnetism'' field theory, which is possibly highly
non-linear. A moving particle is described by a solution of the
theory, such that the ``non-linearity-region'' (or the
``strong-field-region'') is concentrated in a tiny world tube
${\cal W}$ around a smooth, timelike trajectory $\zeta$. We assume
that outside of this tube matter fields practically vanish and the
electromagnetic field is sufficiently weak to be well described by
the linear Maxwell theory. The four momentum of the total system
``particle + field'' is obtained by integration of a (conserved --
due to Noether Theorem) total energy-momentum tensor ${\mathfrak
T}$:
\begin{equation}\label{czteroped1}
  {\cal P}_\lambda  = \int_\Sigma {\mathfrak T}^{\mu}_{\ \lambda}
  d\sigma_\mu \ ,
\end{equation}
over a spacelike hyperplane $\Sigma$.

We assume, moreover, that this fundamental theory admits also a
static, stable, soliton-like solution, which will be called a
``particle at rest''. Here, the strong-field region (interior of
the particle) is assumed to be concentrated around the straight
line ${\vec x}=$const.  Let $m$ denote the {\em total} energy
(mass) of this solution. Due to relativistic invariance, we have
also a six parameter family of solutions obtained by acting with
Poincar\'e transformations on the  static solution. Each of these
solutions may be called a ``uniformly moving particle''. If the
solution has been boosted to the four-velocity $u_{\lambda}$ and
if $\mathsf{T}(u)$ denotes its energy-momentum tensor, then the
total four-momentum of this solution equals $mu_{\lambda}$ and we
have:
\begin{equation}\label{czterroped1}
  mu_{\lambda} = \int_{\Sigma}\mathsf{T}^{\mu}_{\ \lambda}(u)
  d\sigma_\mu \ .
\end{equation}
This leads to a trivial identity:
\begin{equation}\label{cz}
  {\cal P}_\lambda = m u_\lambda +
  \int_{\Sigma} \left( {\frak T}^{\mu}_{\ \lambda} -
  \mathsf{T}^{\mu}_{\ \lambda}(u) \right) d\sigma_\mu \ ,
\end{equation}
which becomes extremely useful in the following arrangement. We
assume that the straight line which describes the ``trajectory''
of the second  (uniformly moving) particle is tangent to the
approximate trajectory $\zeta$ of the first (i.e.~generic)
particle at their intersection point with $\Sigma$. If $K(R)
\subset \Sigma$ denotes the ball of radius $R$, which contains the
strong field region of both solutions, but is small with respect
to the characteristic distance of the external Maxwell fields,
then we have:
\begin{eqnarray}\label{cz1}
  {\cal P}_\lambda &=& m u_\lambda +
  \int_{\Sigma - K(R)} \left( {\frak T}^{\mu}_{\ \lambda} -
  \mathsf{T}^{\mu}_{\ \lambda} (u)\right) d\sigma_\mu +\notag\\
&+&\int_{K(R)} \left( {\frak T}^{\mu}_{\ \lambda} -
  \mathsf{T}^{\mu}_{\ \lambda}(u) \right) d\sigma_\mu
  \ .
\end{eqnarray}
Our assumption about stability of the free particle (soliton
solution) means that the last integral is negligible since inside
the particle both solutions are very close to each other. But the
first integral contains only contributions from external Maxwell
fields accompanying both particles. This way we have proved that
the following formula:
\begin{equation}\label{czteroped2a}
  {\cal P}_\lambda \simeq m u_\lambda +
  \int_{\Sigma - K(R)} \left( {\frak T}^{\mu}_{\ \lambda} -
  \mathsf{T}^{\mu}_{\ \lambda} (u)\right) d\sigma_\mu \ ,
\end{equation}
containing only external Maxwell field surrounding the particle,
provides a good approximation of the total four-momentum of the
total ``particle + field'' system.

The theory proposed in \cite{EMP} consists in mimicking the above
formula in the point particle model. Hence, we consider solutions
of Maxwell equations having a ``delta-like'' current corresponding
to a point charge $e$ traveling over a trajectory $\zeta$. Such a
solution is treated as an idealized description of external
properties of the extended particle considered above. Denote by
$T$ the energy momentum tensor of this solution. Of course, the
uniformly moving particle, whose four-velocity equals $u$, is
represented in this picture by a boosted Coulomb field, and its
energy-momentum tensor is denoted by $\mathbb{T}(u)$. If
trajectories of both particles are again tangent with each other
at their common point of intersection with $\Sigma$, then momentum
(\ref{czteroped2a}) may be rewritten as:
\begin{equation}
  {\cal P}_\lambda \simeq m u_\lambda +
  \int_{\Sigma - K(R)}  \left( { T}^{\mu}_{\ \lambda} -
  \mathbb{T}^{\mu}_{\ \lambda} (u)\right) d\sigma_\mu \ ,
\end{equation}
because outside of the particle ${\frak T}$ reduces to $T$ and
$\mathsf{T}(u)$ reduces to $\mathbb{T}(u)$. The main observation
done in \cite{EMP} is that, due to cancellation of principal
singularities of both $T$ and $\mathbb{T}$(u), the above
integration may be extended to the entire $\Sigma$. More
precisely, the following quantity:
\begin{equation}\label{czteroped3}
  {\cal P}_\lambda := m u_\lambda +
  P \int_{\Sigma} \left( T^{\mu}_{\ \lambda} -
  \mathbb{T}^{\mu}_{\ \lambda} (u)\right) d\sigma_\mu \ ,
\end{equation}
is well defined (``$P$'' denotes the ``principal value'' of the
integral). According to the discussion above, we interpret this
quantity as the total four-momentum of the interacting system
composed of the point particle and the Maxwell field accompanying
the particle. Consequently, we impose conservation of ${\cal P}$
as an additional condition. This implies equations of motion of
the point particle as a good approximation of equations of motion
of the true, extended particle.

This approach has an obvious generalization to the system of many
particles (see \cite{EMP}). Also polarizable particles, carrying
magnetic or electric moment (and -- consequently -- displaying
stronger field singularity than the Coulomb field) may be treated
this way (cf.~\cite{praca-dokt}). Recently, the above approach was
improved by replacing the reference Coulomb field in
(\ref{czteroped3}) by the Born field, matching not only particle's
velocity but also its acceleration. This way the
principal-value-sign ``$P$'' may be omitted in the definition
because the corresponding integral converges absolutely
(cf.~\cite{KP}).

In what follows, we are going to apply definition
(\ref{czteroped3}) to static ``particle + field'' configurations
only.

\section{Electrostatics of a monopole particle}

Consider now electrostatic field $D$ surrounding the particle with
charge $e$, situated at the point $\vec{r}_0$. Due to Maxwell
equations, the Gauss law:
\begin{equation}\label{GL}
  \nabla {D}=e{\boldsymbol \delta}\left(\vec{r}-\vec{r}_0\right) \
  ,
\end{equation}
must be satisfied, where by ${\boldsymbol \delta}$ we denote Dirac
delta distribution (in contrast with conventional $\delta$,
denoting variation of a function). It is, therefore, convenient to
decompose the field into its singular and regular parts:
\begin{equation}
D=\RE{D}+\Si{D},
\end{equation}
where the singular part $\Si{D}$ is simply the Coulomb field:
\[
 \Si{D} := \frac{e \left(\vec{r}-\vec{r}_0\right)}{4\pi \|
  \vec{r}-\vec{r}_0 \|^3} \ ,
\]
whereas the remaining field $\RE{D}:=D - \Si{D}$ is
divergenceless: $\nabla \RE{D} =0$. Moreover, static Maxwell
equations imply the existence of the scalar potential $\phi$:
$D=-\nabla\phi$. Hence, we have: $\Delta\RE{\phi}=0$.

According to (\ref{czteroped3}), the complete energy of this
``particle + field'' system contained in the the entire $\Sigma$
equals:
\begin{equation}\label{energia-calosc}
   {\cal H}=m +
   \frac{1}{2}\int_{\Sigma}\left( {D}^2 - \Si{D}^2
   \right) dv \ .
\end{equation}
We suppose that the particle is contained in a fixed volume $V \ni
\vec{r}_0$. Subtracting from ${\cal H}$ the electrostatic energy
contained outside of $V$:
\begin{equation}\label{energia-outside}
   {\cal H}_{{\mathbb R^3} - V}=
   \frac{1}{2}\int_{{\mathbb R^3} - V} {D}^2 dv \ ,
\end{equation}
we obtain the total energy contained in $V$:
\begin{eqnarray}\label{calka-zren-energia}
{\cal H}_{V}&=&m  -\frac{1}{2}\int_{{\mathbb R^3}-V}{\Si{D}}^2
dv+\frac{1}{2}\int_{V} {\RE{D}}^2 dv+\notag\\
&+&\int_{V}\Si{D}\RE{D}dv.
\end{eqnarray}
Given boundary conditions, we are going to minimize the above
quantity with respect to the particle's position $\vec{r}_0 \in
V$. Assuming that the particle always tries to minimize the energy
of the system, we can write both $\vec{r}_0$ and the total
``particle+field'' energy as functions of the field boundary data.
Stability of the energy with respect to the boundary data on
$\partial V$ will then be studied. Before we pass to the above
programme, we must specify which kind of boundary conditions on
$\partial V$ have to be controlled.
\subsection{Neumann conditions}
Varying the energy integral (\ref{calka-zren-energia}) with
respect to the particle's position we get:
\begin{align}\label{calka:energii}
\delta{\cal
H}_V=&\int_V\left\{\RE{D}\cdot\left(\delta\RE{D}+
\delta\Si{D}\right)+
\Si{D}\,\delta\RE{D}\right\}dv \notag\\ &-\int_{{\mathbb
R}^3-V}\Si{D}\,\delta\RE{D}\,dv.
\end{align}
For Neumann conditions we put $D=-\nabla\phi$ for both the regular
and the singular parts of the field, outside of the variation
$\delta$. Integrating by parts and using $\nabla\RE{D}=0$ we get:
\begin{equation}\label{3.7}
\delta{\cal
H}_V=\int_V\RE{\phi}\,\delta(\nabla\Si{D})dv-\int_{\partial
V}\left\{ \phi\,\delta D^{\bot}\right\}\,d\sigma.
\end{equation}
But the variation of (\ref{GL}) gives us:
\begin{equation}\label{wariacja:dsing:mono}
\delta(\nabla\Si{D})=\delta\left(e{\boldsymbol
\delta}\left(\vec{r}-\vec{r}_0\right)\right)=-e\partial_k
\left({\boldsymbol\delta}\left(\vec{r}-\vec{r}_0\right)\right)\delta
x_0^k,
\end{equation}
where $\delta x_0^k$ denotes a virtual displacement of the
particle. Imposing Neumann conditions $D^{\bot}|\partial V = f$,
where $f$ is a fixed function, we obtain: $\delta D^{\bot} \equiv
0$ on $\partial V$. Hence, the surface integral vanishes.
Inserting (\ref{wariacja:dsing:mono}) into (\ref{3.7}) we derive
the following formula:
\begin{equation}\label{neumann}
\delta {\cal H}_V=-e D^{\text{reg}}_k( x^k_0)\delta x^k_0.
\end{equation}
We conclude that the extremum of  energy condition implies the
following static equilibrium equation:
\begin{equation}\label{equilibrium}
  D^{\text{reg}}_k ( x^k_0) = 0 \ .
\end{equation}

\subsection{Dirichlet conditions}
For Dirichlet case  we put $\delta D=-\nabla\delta\phi$ for both
the regular and the singular parts of the field  and then
integrate (\ref{calka:energii}) by parts. We obtain:
\begin{equation}
 \delta{\cal H}_V=\int_V \left(\nabla\Si{D}\right)\delta\RE{\phi}dv
 - \int_{\partial V}\left\{
 D^{\bot}\,\delta\phi \right\}\,d\sigma.
\end{equation}
Imposing Dirichlet conditions $\phi|\partial V = f$, where $f$ is
a fixed function, we obtain: $\delta \phi \equiv 0$ on $\partial
V$ and, therefore, the surface integral vanishes again. To derive
the equilibrium condition (\ref{equilibrium}) from the variational
principle, we must perform the following Legendre transformation:
\begin{align}\label{Legendre1}
  \int_V \left(\nabla\Si{D}\right)\delta\RE{\phi}dv &=
  \delta\int_V \left(\nabla\Si{D}\right)\RE{\phi}dv+\nonumber\\
 &-\int_V \left(\delta\nabla\Si{D}\right)\RE{\phi}dv \ .
\end{align}
Then we use (\ref{GL}) and (\ref{wariacja:dsing:mono}).
This way we obtain:
\begin{equation}\label{dirichlet}
 \delta\left(
 {\cal H}_V-e\RE{\phi}({\vec{r}}_0)\right)=
 D^{\text{reg}}_k( x^k_0)\delta x^k_0.
\end{equation}
Comparing (\ref{neumann}) and (\ref{dirichlet}) we observe that
the equilibrium condition (\ref{equilibrium}) may either be
obtained from the variational principle $\delta\left( {\cal
H}_V\right) =0$, when the Neumann boundary data are controlled, or
from the variational principle $\delta\left( {\cal F}_V\right)
=0$, with ${\cal F}_V := {\cal H}_V-e\RE{\phi}({\vec{r}}_0)$, when
the Dirichlet boundary data are controlled. The quantity ${\cal
H}_V$ is the total energy of the ``particle + field'' system,
whereas ${\cal F}_V$ is an analog of the free energy in
thermodynamics. We conclude that imposing  Neumann condition on
the boundary corresponds to the adiabatic insulation of the
system, whereas imposing Dirichlet condition means that we expose
it to a kind of a ``thermal bath''. Indeed, imposing
e.g.~condition $\phi|\partial V =0$ we must cover the surface
$\partial V$ with a metal shell and ground it electrically. This
means that we admit energy exchange of our system with the earth.
Similarly as in thermodynamics, the free energy ${\cal F}_V$,
which we optimize, contains not only the system's energy ${\cal
H}_V$ but also the term ``$-e\RE{\phi}({\vec{r}}_0)$'' which we
interpret as energy of the ``boundary-condition-controlling
device''. Of course, from the point of view of the particle, both
conditions lead to the same equation: $D^{\text{reg}}( x^k_0) = 0$
because our theory is \emph{local} and the particle interacts with
its immediate neighbourhood only, no matter how the boundary data
are controlled far away from the particle.
\section{An example  -- monopole particle in a spherical box}
\label{mon-part-sph-box} In this section we shall analyze
stability of a charged, monopole particle closed in a spherical
box with radius $R$: $V = K(0,R) \subset {\mathbb R}^3$.
Simplicity of the model allows us to solve explicitly the static
Maxwell equations (for both the Neumann and the Dirichlet cases)
and to compute renormalized energy of the system. Then we will
find the {\em extremum} of the energy function with respect to the
particle's position and check that for the Neumann case we get the
minimum and for the Dirichlet case --  the maximum of the energy.
Assuming that the particle always minimizes the energy, we will
express energy function in terms of the boundary data and show
that the system is unstable under small changes of these data.

The problem consists in solving equation $\Delta\phi =-
e\delta(\vec{r}-\vec{r}_0)$, where $\vec{r}_0 \in K(0,R)$. In the
Neumann case we impose the following condition:
\begin{align}\label{Neu-co}
\vec{D}\cdot\vec{n}{\big\vert}_{r=R}=\vec{E}\cdot\vec{n}
+\frac{e}{4\pi R^2} \ ,
\end{align}
where $\vec{E}$ is a fixed three dimensional vector.

In the Dirichlet case we impose the following condition:
\begin{align}\label{Dir-co}
\phi{\big\vert}_{r=R}=-\vec{E}\cdot\vec{n}\,R+\frac{e}{4\pi R} \ .
\end{align}
Because of the axial symmetry of the problem, we may restrict
ourselves to the analysis of the energy functional at points
$\vec{r}_0$ which are parallel to $\vec{E}$: $\vec{r}_0 \|
\vec{E}$.
 With this simplification, we are able to
find an explicit solution $\phi=\Si{\phi}+\RE{\phi}$, where:
\[
 \Si{\phi}=\frac{1}{4\pi}\frac{e}{|\vec{r}-\vec{r_0}|} \ ,
\]
in both Dirichlet and Neumann cases (cf.~Appendices
\ref{dodatek:pole:elektryczne} and
\ref{dodatek:pole:elektryczne:dirichlet}). To write an explicit
formula for $\RE{\phi}$ it is useful to introduce the following
variable:
\[
 r_0 := \frac 1{\| E \|} (\vec{E} | \vec{r}_0 )  \  ,
\]
which runs from $-R$ to $R$. Under this convention we obtain:
\begin{align}
&\RE{\phi}=\frac{e}{4\pi}\bigg(\frac{R}{\sqrt{R^4+{r_0}^2 r^2-2
r_0 r R^2\cos\theta}} -\frac{1}{R}\nonumber+\\
& -\frac{1}{R}\ln\left|R^2-r_0 r\cos\theta+ \sqrt{R^4+{r_0}^2
r^2-2 r_0 r R^2\cos\theta}\right|\bigg)\nonumber\\
&- \vec{E}{\vec{r}}+\frac{1}{R}\ln(2 R^2) \ ,
\label{phir}
\end{align}
in the Neumann case, whereas:
\begin{align}
\RE{\phi}&=\frac{e}{4\pi}\bigg(\frac{1}{R}-
\frac{R}{\sqrt{R^4+{r_0}^2 r^2-2 r_0 r R^2\cos\theta}}\bigg) -
\vec{E}{\vec{r}} \ , \label{phird}
\end{align}
in the Dirichlet case.
\subsection{Stability}
\begin{figure}
\centering\epsfig{file=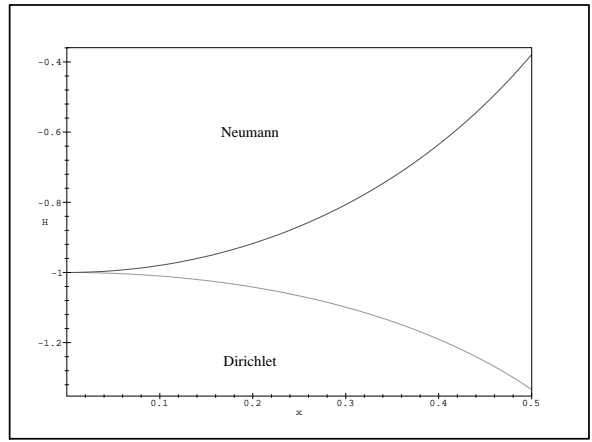,height=8.5cm, angle=-90}
\caption{Graph of renormalized energy vs particle's
position and $q=0$ for  ${\cal {H'}_N}$
 and ${\cal {H'}_D}$}\label{rysunek:neu:diri}
\end{figure}

In both cases, the renormalized energy can be computed explicitly.
Denoting $E:=\| \vec{E} \|$ we obtain the following result:
\begin{align}\label{energia:nneeumann}
{\cal H}_{\cal N}=&m+\frac{1}{2}\left(\frac{e^2}{4\pi}
\left(\frac{R}{R^2-r_0^2}
-\frac{1}{R}\ln\left|1-\frac{r_0^2}{R^2}\right|-\frac{2}{R}\right)+\right.\nonumber\\
&\left.+\frac{4}{3}\pi R^3 E^2-2 e E r_0\right),
\end{align}
in the Neumann case (cf.~Appendix \ref{dodatek:energia}) and:
\begin{equation}\label{energia:dirichleta}
{\cal H}_{\cal D}=m+\frac{1}{2}\left(\frac{4}{3}\pi R^3
E^2-\frac{e^2}{4\pi} \frac{R}{R^2-r_0^2}\right) \ ,
\end{equation}
in the Dirichlet case (cf.~Appendix
\ref{dodatek:pole:elektryczne:dirichlet}). Finally, we compute the
electric ``free energy'' ${\cal F}={\cal H}
-e\RE{\phi}(\vec{r}_0)$ in the Dirichlet case:
\begin{equation}
{\cal F}=m+\frac{1}{2}\left(\frac{e^2}{4\pi}\frac{R}{R^2-r_0^2}
+2e E r_0+\frac{4}{3}\pi R^3 E^2
-\frac{e^2}{4\pi}\frac{2}{R}\right).
\end{equation}
We see that the equilibrium condition in the Neumann case reads:
\begin{gather}\label{eqi-N}
\RE{D}{\big\vert}_{\vec{r}=\vec{r}_0}=0 \Leftrightarrow
\left(eE-\frac{e^2}{4\pi}\frac{r_0}{R(R^2-r_0^2)}\right)=0 \ ,
\end{gather}
whereas in the Dirichlet case it reads:
\begin{gather}
e\RE{D}{\big\vert}_{\vec{r}=\vec{r}_0}=\frac{e^2}{4\pi}\frac{R
r_0}{(R^2-r_0^2)^2}+e E = \frac{\partial}{\partial r_0} {\cal F} \
.
\end{gather}
We express the energy in terms of the following, standardized
variables:
\begin{equation}\label{standard}
 x=\frac{r_0}{R}\in ]-1,1[,\qquad  q=\frac{4\pi R^2}{e}
 E \ .
\end{equation}
Denoting:
\begin{equation}\label{H-standard}
  {\cal H'}= ({\cal H}-m)\frac{8\pi R}{e^2} \ ,
\end{equation}
we  obtain:
\begin{align}
{\cal {H'}_N}&=\frac{1}{1-x^2}-\ln|1-x^2|-2 q x +\frac{1}{3} q^2 -2,\\
{\cal {H'}_D}&=\frac{1}{3} q^2-\frac{1}{1-x^2}.
\end{align}
Observe that for $q=0$ both energies may be expanded as follows
(cf.~figure \ref{rysunek:neu:diri}):
\begin{eqnarray}\label{energia:monopola}
 {\cal {H'}_N}&=& -1 + 2 x ^2 + O(x^4) \ ,
 \\
 {\cal {H'}_D}&=&  -1-x^2
 + O(x^4) \ .
\end{eqnarray}
This implies that only in the Neumann case the equilibrium point
($x=0$) is also a minimum of the energy. In the Dirichlet case the
energy has a local maximum at the equilibrium point. As may be
easily seen, this happens also for any value of $E$. Hence, for
the Dirichlet case the free energy $\cal F$ should be used, for
which local extremum is also minimum. In what follows we shall use
the local, physical energy and consequently, we restrict ourselves
to the Neumann case only.
\subsection{Neumann conditions}
\begin{figure}
\centering
\epsfig{file=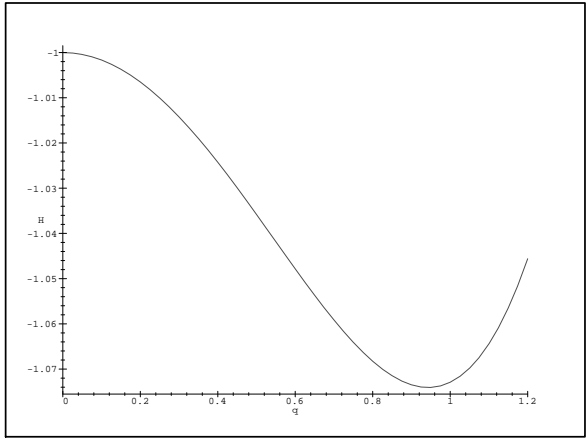,height=8.5cm, angle=-90}
\caption{Graph of renormalized energy vs boundary
field $q$ for ${\cal {H'}_N}(x(q),q)$}
\label{rysunek:energia:mono}
\end{figure}

In terms of the standardized variables, the equilibrium condition
(\ref{eqi-N}) reads:
\begin{equation}
   q = \frac{x(2-x^2)}{(1-x^2)^2} \ .
\end{equation}
For small values of $q$ this enables us to express equilibrium
position in terms of the boundary data:
\begin{equation}\label{q-approx}
  x\approx
\frac{q}{2} \ .
\end{equation}
The same result could be obtained from the following expansion:
\begin{gather}
 {\cal {H'}_N}(x,q)= -1+\frac{1}{3} q^2-2q x +2 x^2+ O(x^4), \\
 \partial_{x}{\cal {H'}_N}(x,q)=0 \Rightarrow x\approx \frac{q}{2},
 \label{hq:neumann}\\ {\cal {H'}_N}(x,q){\vert}_{x=\frac{q}{2}}=
 -1-\frac{1}{6}q^2   + O(q^3).
\end{gather}
Observe that for \emph{increasing} values of $q$,  the energy of
the system \emph{decreases} (cf.~figure
\ref{rysunek:energia:mono})! The system ``particle + field'' turns
out to be \emph{unstable} -- even small fluctuations of the
external field $q$ can decrease its total energy.  This means that
the particle behaves like a {\em perpetuum mobile}, providing a
source of energy at no costs. In our opinion this unphysical
feature of the model, manifestly seen in its static behaviour,
could possibly be a source of its dynamical instability, i.e.~the
existence of ``runaway'' solutions of Dirac equation. As a remedy,
described in the sequel, we propose to equip the particle with an
additional mechanism which, {\em via} electric polarizability,
will restore its static stability.

\section{Polarizable particle}
\nopagebreak We assume that the particle may get a non-vanishing
electric dipole moment due to interaction with the neighboring
field. We prove in the sequel that, under a suitable choice of the
polarizability properties of the particle, the resulting
``particle + field'' system becomes statically stable.

For a polarized particle, formula (\ref{calka-zren-energia}) for
the total energy remains valid but the field singularity is now
deeper than in (\ref{GL}), namely:
\begin{align}
   & \nabla D =
    \nabla\Si{D}=e{\boldsymbol\delta}(\vec{r}-\vec{r}_0)-
    p^k\partial_k{\boldsymbol\delta}(\vec{r}-\vec{r}_0) \ ,\label{rownanie-z-dipolem}
\end{align}
where $p^k$ is a dipole moment. We assume that $p^k$ has been
generated by the surrounding electric field $D$ according to some
law $p=p(D_{reg}(\vec{r}_0))$, describing the sensitivity of the
particle. Moreover, we admit the dependence of the coefficient $m$
in (\ref{czteroped3}) (and, consequently, in
(\ref{calka-zren-energia})) upon polarization. It will be shown in
the sequel that insisting in having $m$ constant we are not able
to make the model physically consistent. Moreover, it will be
shown that the electric sensitivity is {\em uniquely} implied by
the dependence $m=m(p)$.
\subsection{Variational  principle}
Variation of the renormalized energy (\ref{calka-zren-energia})
with respect to the particle's position contains now the
non-vanishing term $\delta m$. Similar calculations as for the
scalar particle lead, in case of the Neumann boundary conditions,
to formula:
\begin{eqnarray}
  \delta{\cal H}_V&=&\delta m+\int_{V}\RE{\phi}\delta(\nabla\Si{D})
  dv +\nonumber\\
&-&\int_{\partial V}\left\{
  \phi\, \delta D^\perp \right\}\,d\sigma
  \label{war-Neu-dip} \ ,
\end{eqnarray}
and, in case of the Dirichlet conditions, to:
\begin{align}
  & \delta {\cal H}_V
     =  \delta m + \int_V  (\nabla \Si{D}) \delta  \RE{\phi} dv -
     \int_{\partial V}\left\{ D^\perp \delta \phi
    \right\}\,d\sigma =\nonumber\\
& \delta m
    +  \delta \int_V  (\nabla \Si{D}) \RE{\phi} dv  -
    \int_{V}\RE{\phi}\delta(\nabla\Si{D}) dv+\nonumber \\
&-\int_{\partial V}\left\{ D^\perp \delta \phi
    \right\}\,d\sigma
      \ . \label{war-Dir-dip}
\end{align}
According to (\ref{rownanie-z-dipolem}), the new version of
formula (\ref{wariacja:dsing:mono}) reads:
\begin{align}\label{delta-ladunku}
  \delta (\nabla \Si{D}) & =  - \left( e \partial_k {\boldsymbol\delta}
  (\vec{r}-\vec{r}_0) -
  p^j\partial_j\partial_k{\boldsymbol\delta}(\vec{r}-\vec{r}_0)
  \right)\delta  x^k_0 +\nonumber\\
&-  \left(\partial_k{\boldsymbol\delta}
  (\vec{r}-\vec{r}_0)\right)  \delta   p^k \ .
\end{align}
Plugging (\ref{delta-ladunku}) into  (\ref{war-Neu-dip}) we see
that the total energy variation splits into the sum of two pieces:
the work due to virtual displacement of the particle and the
remaining work, due to variation of $m$ and $p$:
\begin{align}\label{war-H-dip}
  \delta{\cal  H}_V&=\underbrace{-\left(e\RE{D}+p^k
  \partial_k\RE{D} \right){\big\vert}_{\vec{r}=
  \vec{r}_0}}_{{\cal A}}\delta\vec{r}_0 +\nonumber\\
&+  \underbrace{\delta m - \RE{D}{\big\vert}_{\vec{r}=
  \vec{r}_0}\delta p}_{{\cal B}} \ .
\end{align}
The second part $\cal B$ is obviously  \emph{nonlocal} -- both the
mass $m$ and the moment $p$ depend upon the value of
$\RE{D}(\vec{r}_0)$. This quantity must be obtained from the field
equation: $\Delta \RE{\phi}=0$, with boundary value depending upon
the particle's position. The only way to save locality of the
model is to force the term $\cal B$ to vanish \emph{identically}
by imposing the following constraint:
\begin{equation}\label{gener}
  \delta m = \RE{D}(\vec{r}_0)\delta p \ .
\end{equation}
Denoting by $m_0=m(0)$ the mass of the unpolarized particle and by
$f(p)$ the additional polarization energy:
\begin{equation}\label{m-0}
  m(p) = m_0 + f(p) \ ,
\end{equation}
formula (\ref{gener}) may be written as:
\begin{equation}\label{gener1}
  D^{reg}_k(\vec{r}_0) = \frac {\partial f(p)}{\partial p^k} \ .
\end{equation}
We see that the polarization energy $f$ must play role of the
generating function for the polarizability relation, otherwise the
model would not be local. Indeed, suppose that ${\cal B}$ does not
vanish and the particle's equilibrium condition needs vanishing of
the whole right hand side of (\ref{war-H-dip}). To decide whether
or not its actual position is acceptable as an equilibrium
position, the particle must know not only the field in its
immediate neighbourhood, but also the shape of $V$ and the field
boundary data on $\partial V$. Such a behaviour is physically non
acceptable.

Inverting the generating formula (\ref{gener1}), we may find the
dependence $p = p(\RE{D}(\vec{r}_0) )$, which is uniquely implied
by the ``equation of state'' (\ref{m-0}). Hence, we have:
\begin{equation}\label{war-H-dip1}
  \delta{\cal  H}_V=-\left(e\RE{D}+p^k
  \partial_k\RE{D} \right){\big\vert}_{\vec{r}=
  \vec{r}_0}\delta\vec{r}_0 \ ,
\end{equation}
and the equilibrium condition becomes a local equations:
\begin{equation}\label{rownowaga-dip}
  \left(e\RE{D}+p^k
  \partial_k\RE{D} \right){\big\vert}_{\vec{r}=
  \vec{r}_0} = 0 \ .
\end{equation}

A similar procedure works in the Dirichlet case as well. Applying
the state equation to (\ref{war-Dir-dip}) we obtain:
\begin{equation}
  \delta {\cal F}_V =\left(e\RE{D}+p^k
  \partial_k\RE{D} \right){\big\vert}_{\vec{r}=
  \vec{r}_0}\delta\vec{r}_0 \ ,
    \end{equation}
where the ``free energy'' ${\cal F}_V$ is given as:
\begin{align}
  & {\cal F}_V := {\cal H}_V - \int_V  (\nabla \Si{D})
   \RE{\phi}- 2f \nonumber \\
   &:=  {\cal H}_V - e \phi_{reg}(\vec{r}_0) +
    \RE{D}{\big\vert}_{\vec{r}=
  \vec{r}_0}\cdot p - 2f \ . \label{F-dip}
\end{align}
Equilibrium condition $\delta {\cal F}_V=0$ reduces to the same,
local equation (\ref{rownowaga-dip}).
\section{An example -- polarizable particle in a spherical box}
Let us come back to the simple model described in Section
\ref{mon-part-sph-box} on page \pageref{mon-part-sph-box}. For the
polarizable particle we must solve the field equation:
\begin{align}
\Delta\phi =- e\delta(\vec{r}-\vec{r}_0)+\vec{p}\cdot\nabla(
\delta(\vec{r}-\vec{r}_0)) \ ,
\end{align}
where $\vec{r}_0 \in K(0,R)$, with either Neumann (\ref{Neu-co})
or Dirichlet condition (\ref{Dir-co}). We want to compute
renormalized total energy of the ``particle + field'' system and
to prove that for a suitable state equation (\ref{m-0}) our model
becomes stable.

Splitting the solution $\phi$ into two parts:
\begin{equation}
  \phi =\phi^{mon}+\phi^{dip} \ ,
\end{equation}
where by $\phi^{mon}$ we denote the solution of the monopole
problem, found earlier (cf. Section (\ref{mon-part-sph-box}), page
\pageref{phir}), we reduce the problem to equation:
\begin{equation}
  \Delta\phi^{dip} = \vec{p}\cdot\nabla\,(\delta(\vec{r}-\vec{r}_0))
  \  ,
\end{equation}
with homogeneous boundary conditions:
$\vec{D}^{dip}\cdot\vec{n}{\big\vert}_{r=R}=0$ in the Neumann case
and $\phi^{dip}{\big\vert}_{r=R}=0$ in the Dirichlet case.
Choosing the axis ${\boldsymbol e_z}$ parallel to $\vec{E}$ and
passing to spherical coordinates $(r,\theta, \varphi)$ we obtain
for $\vec{r}_0=(r_0,0,0)$ and $\vec{p}=p{\boldsymbol
e_z}+p_x{\boldsymbol e_x}$ (see Appendix \ref{dodatek:pole:dipola}
on page \pageref{dodatek:pole:dipola}):
\begin{equation}
 \phi^{dip}=\Si{\phi^{dip}}+\RE{\phi^{dip}} \ ,
\end{equation}
where:
\begin{align}
&\Si{\phi^{dip}} =
\frac{1}{4\pi}\frac{\vec{p}\cdot(\vec{r}-\vec{r}_0)}{|\vec{r}
-\vec{r}_0|^3},\\
&\RE{\phi^{dip}}=\frac{p}{4\pi}\left(\frac{R^3\left(R^2-r
r_0\cos\theta\right)}{r_0 \left(R^4+(r_0 r)^2-2r r_0
R^2\cos\theta\right)^{\frac{3}{2}}} -\frac{1}{r_0
R}\right)\nonumber\\& +\frac{p_x}{4\pi}\Bigg(\frac{r
R^3\sin\theta\cos\varphi} {(R^4+{r_0}^2 r^2-2 r_0 r
R^2\cos\theta)^{\frac{3}{2}}}\notag+\\
&-\frac{\cos\varphi(R^2\cos\theta-r_0 r)}{R r_0\sin\theta
\sqrt{R^4+{r_0}^2 r^2-2 r_0  r
R^2\cos\theta}}+\frac{\cos\theta\cos\varphi} {R
r_0\sin\theta}\Bigg). \label{phir-dip}
\end{align}
As we already noticed in the monopole case, axial symmetry of the
problem implies that minimum of the energy is assumed at the point
$\vec{r}_0$ which is parallel to $\vec{E}$. The same argument
implies that we have ${p_x}=0$ in this configuration. We are going
to limit our analysis to such configurations only.

\subsection{Stability}
We compute the total, renormalized energy of the system as a sum
of two parts:
\begin{equation}
  {\cal H} ={\cal H}^{mon}+{\cal H}^{dip} \ ,
\end{equation}
where ${\cal H}^{mon}$ denotes the energy of the monopole field
obtained earlier ((\ref{energia:nneeumann}), page
\pageref{energia:nneeumann}), and  ${\cal H}^{dip}$ denotes the
remaining part, containing energy of the dipole field and the
interaction energy. The latter term is computed in Appendix
\ref{dodatek:energia:dipolowa} (page
\pageref{dodatek:energia:dipolowa}). The final result for the
Neumann case, written in terms of standardized variables reads:
\begin{align}\label{energia:Hprim}
 & {\cal H_{N}}'(x,q,p)=\frac{1}{1-x^2}-\ln|1-x^2|-2 q x
  +\frac{1}{3} q^2 -2+
  \nonumber \\
  &+
 \frac{2}{3}\left(\frac{p}{ e R}\frac{x(2-x^2)}{(1-x^2)^2}
 -\frac{p^2}{e^2 R^2}\frac{1}{(1-x^2)^3}-\frac{p}{e R}\,q\right).
\end{align}
Now, stability of the system depends upon the polarizability of
the particle, i.e.~upon the choice of the ``state function'' $f$
(cf.~(\ref{m-0}) on page \pageref{m-0}). At the moment we have no
general criterion which would guarantee stability. However, it is
easy to show that for:
\begin{equation}\label{pierwiastek}
f(\vec{p})=-\frac{c^2}{3}\|\vec{p}\|^3 \Longrightarrow
\RE{D}=-c^2\|\vec{p}\|\vec{p},\quad c>0 \ ,
\end{equation}
our system is stable. Indeed, using (\ref{phir}) and
(\ref{phir-dip}) we obtain the following equation for the value of
the dipole moment $p$:
\begin{align}
  &-c^2p^2 \mbox{\rm sgn} ( p )=
  \RE{D}{\bigg\vert}_{\vec{r}=\vec{r}_0} =
  -\nabla\left(\RE{{\phi}^{mon}}+\RE{{\phi}^{dip}}\right)=\nonumber\\
   &= \frac1{4\pi} \left(\frac{e q}{R^2}-\frac{2
  p}{R^3(1-x^2)^3}-
  \frac{e x(2-x^2)}{R^2(1-x^2)^2}\right) \ .
\end{align}
Denoting $4\pi e c^2 R^4 = C$ and ${\widetilde p} = \frac p{eR}$,
we get equation for $\widetilde p$:
\begin{equation}\label{Cp}
   -C  {\widetilde p}^2 \mbox{\rm sgn} ( {\widetilde p} )=
   \left(q-\frac{2 {\widetilde p}}{ (1-x^2)^3}-
  \frac{ x(2-x^2)}{(1-x^2)^2}\right)\ .
\end{equation}
For small $x$, we use Taylor expansion of the right hand side.
Consequently, we have:
\begin{equation}
  -C{\widetilde p}^2 \mbox{\rm sgn} ({\widetilde p} )
   \approx q- 2{\widetilde p}-2x-6{\widetilde p}x^2-3x^3.
\end{equation}
For ${\widetilde p}>0$ there are two solutions of this equation
for small $x$ and $q$:
\begin{eqnarray}\label{p1}
  {\widetilde p}_1 & \approx & \frac{1}{C}\left(1+\sqrt{1-qC}
+ \frac{x C}{\sqrt{1-qC}}\right) \ , \\
 {\widetilde p}_2 & \approx & \frac{1}{C}\left(1-\sqrt{1-qC}
 -\frac{x C}{\sqrt{1-qC}}\right) \ . \label{p2}
\end{eqnarray}
For ${\widetilde p}<0$ there is only one solution for small $x$
and $q$:
\begin{equation}
 {\widetilde p}_3 \approx -\frac{1}{C}\left(1+\sqrt{1+qC}
 -\frac{x C}{\sqrt{1+qC}}\right).
\end{equation}
Inserting the above solutions into the energy function
(\ref{energia:Hprim}) we define for $i=1,2,3$:
\[
 {\cal H'}_i(x,q) = {\cal
  H_{N}}'(x,q,eR{\widetilde p}_i) \ .
\]
It turns out that ${\cal H'}_2$ does not admit any minimum with
respect to $x$ (i.e.~a stable ``field + particle'' configuration).
For the remaining two cases we use Taylor expansion for small $x$:
\begin{align}
& {\cal H'}_1
 \approx -1+\frac{1}{3} q^2 -\frac{4}{3C^2}+\frac{2}{3\sqrt{1-qC}}
 \left(-\frac{2}{C^2}+\frac{q}{C}+q^2\right) \notag\\
 &- 2q\left( 1+\frac{1}{\sqrt{1-qC}}\right)x
 +2\left(1-\frac{2}{C^2}+\frac{q}{C}-\frac{1}{3}\frac{1}{1-qC}
 \right.\notag\\
&\left. +\frac{2}{\sqrt{1-qC}}\left(\frac{1}{3}
 -\frac{1}{C^2}+\frac{q}{C}\right)\right)x^2.
\end{align}
\begin{align}
& {\cal H'}_3
  \approx -1+\frac{1}{3} q^2 -\frac{4}{3C^2}+\frac{2}{3\sqrt{1+qC}}
 \left(-\frac{2}{C^2}-\frac{q}{C}+q^2\right) \notag\\
 &- 2q\left( 1+\frac{1}{\sqrt{1+qC}}\right)x
 +2\left(1-\frac{2}{C^2}-\frac{q}{C}-\frac{1}{3}\frac{1}{1+qC}
 \right.\notag\\
&\left. +\frac{2}{\sqrt{1+qC}}\left(\frac{1}{3}
 -\frac{1}{C^2}-\frac{q}{C}\right)\right)x^2\\
\end{align}

\begin{figure}
\centering \epsfig{file=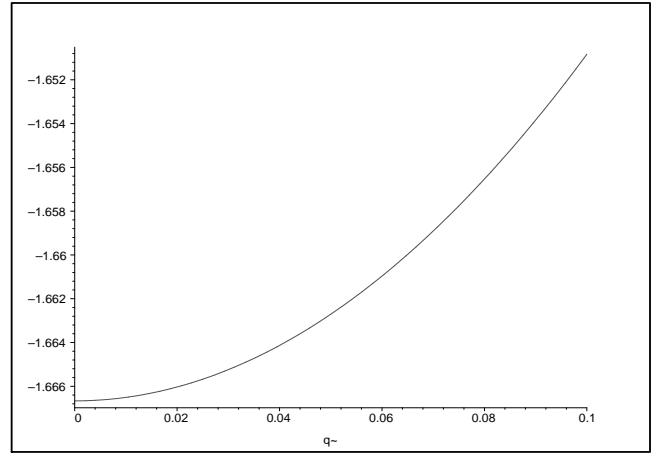,height=8.5cm,width=6cm,
angle=-90} \caption{Graph of  ${\cal H'}(q)$ --  renormalized
energy vs  boundary field for dipole particle, $C=1$}
\label{rysunek:energia:dipo}
\end{figure}

Minimizing both energies with respect to $x$ we obtain:
\begin{align}
 x_1(q)\approx &\frac{3C^2}{32(C^2-3)}\left(8q+\frac{2C(C^2-9)}
 {C^2-3}q^2+\right.\nonumber\\
+&\left.\frac{C^2(2C^4-15C^2+45)}{(C^2-3)^2}q^3\right),\\
 x_3(q)\approx &\frac{3C^2}{32(C^2-3)}\left(8q-\frac{2C(C^2-9)}
 {C^2-3}q^2+\right.\nonumber\\
&+\left.\frac{C^2(2C^4-15C^2+45)}{(C^2-3)^2}q^3\right).
\end{align}
Plugging $x_i(q)$ into the energy we get for small $q$:
\begin{align}
  {\cal H'}_1(q)&\approx
  -1-\frac{8}{3C^2}+\frac{15+4C^2}{6(3-C^2)}q^2+\nonumber\\
& +\frac{(18+42C^2-7C^4)C}{12(3-C^2)^2}q^3,\\
   {\cal H'}_3(q)&\approx -1-\frac{8}{3C^2}+\frac{15+4C^2}{6(3-C^2)}q^2+\nonumber\\
& -\frac{(18+42C^2-7C^4)C}{12(3-C^2)^2}q^3 .
\end{align}
We see that for $C\in]0,\sqrt{3}[$ the $q^2$ term is positive.
This means that the system ``particle + field'' does not behave
any longer like a {\em perpetuum mobile}: to deform its original
configuration, corresponding to $q=0$, the boundary-condition
controlling device must perform a positive work. Hence, the system
is \emph{stable} under small changes of $q$ (see figure
\ref{rysunek:energia:dipo}).

\subsection{Conclusions}
We have shown that the polarizability of the particle, described
by a suitable ``state function'' $f$ (e.g.~by
(\ref{pierwiastek})), may be a good remedy for the static
instability of the renormalized electrodynamics of point
particles. Whether or not this will cure also the dynamical
instability, i.e.~the existence of ``runaway'' solutions, is
another question which we would like to study in the nearest
future.

At the moment the bifurcation phenomenon occurring near the ground
state $q=0$ is worthwhile to study. Observe that the point
$\vec{r}_0 =0$, corresponding to $q=0$ and described by the
purely monopole field, {\em is not} stable. This configuration
corresponds to a local maximum of the energy and belongs to the
unstable branch of stationary points, described by the function
${\cal H'}_2$.


\appendix
\section*{Appendices}
\section{Neumann solution for ``particle + field system''}
\label{dodatek:pole:elektryczne}
\begin{figure}
\centering \epsfig{file=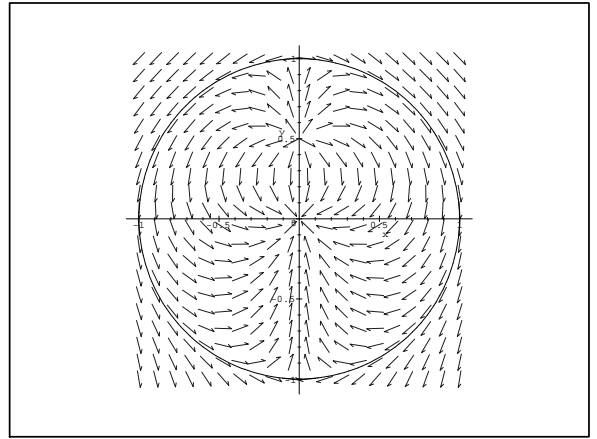,height=8.5cm, angle=-90}
\caption{Directions of the field $D - E$ for $R=1$, $r_0=0.5$}
\label{rysunek:pole:monopola}
\end{figure}

 We are looking for a solution of the Poisson equation
 $\Delta\phi =- e\delta(\vec{r}-\vec{r}_0)$ with boundary condition
({\ref{Neu-co}}), where $\| \vec{r}_0 \| < R$ and $\vec{r}_0 \|
\vec{E}$. Denote:
\begin{equation}\label{phi-nowe}
  \phi=\Si{\phi}+\overbrace{\RE{\phi}^{0}-\vec{E}\vec{r}}^{\RE{\phi}}
  \ ,
\end{equation}
where $\Si{\phi} = \frac{1}{4\pi}\frac{e}{|\vec{r}-\vec{r}_0|}$,
$\Delta\RE{\phi}^{0} =0$ and:
\begin{align}
\RE{\vec{D}}^{0}\cdot\vec{n}{\big\vert}_{r=R}
=\vec{n}\cdot\frac{1}{4\pi}\nabla
\left(\frac{e}{|\vec{r}-\vec{r}_0|}\right)
{\bigg\vert}_{r=R}+\frac{e}{4\pi R^2} \ .
\end{align}
To find $\RE{\phi}^{0}$, we use the following formula
(cf.~\cite{Panofsky}, p.83):
\begin{equation}
 \frac{1}{\sqrt{r^2+{r_0}^2-2\,r r_0 \cos\theta}}-\frac{1}{r}=
 \sum_{n=1}^{\infty}\label{A4}
 P_n(\cos\theta)\frac{{r_0}^n}{r^{n+1}},
\end{equation}
($\theta$ is the angle between $\vec{r}$ and $\vec{E}$) valid for
$-r\leq r_0\leq r$, together with the following {\em Ansatz}:
\begin{equation}
\RE{\phi^{0}}=\sum_{n=1}^{\infty} c_n r^n P_n(\cos\theta)\label{A5}.
\end{equation}
Write boundary condition as:
\begin{equation}
\frac{\partial}{\partial
r}\RE{\phi}^{0}{\bigg\vert}_{r=R}=\frac{e}{4\pi}\frac{\partial}{\partial
r} \left(\frac{1}{r}-\frac{1}{|\vec{r}-\vec{r}_0|}\right)
{\bigg\vert}_{r=R} \ ,\label{A6}
\end{equation}
and substitute (\ref{A4}) and (\ref{A5}) to (\ref{A6}). This way
we get the solution given as a series:
\begin{equation}\label{A7}
\RE{\phi}^{0}=\frac{e R}{4\pi} \sum_{n=1}^{\infty}
 \left(1+\frac{1}{n}\right) \frac{{(r r_0)}^n}{{(R^2)}^{n+1}}
 P_n(\cos\theta) \ .
\end{equation}
Observe that (\ref{A4}) gives, after rescaling, the first
component of (\ref{A7}). The second one will be obtained from the
following:
\begin{lemma}\label{FA1}
For $\| r_0\| \geq r$ we have
\begin{align}
& \sum_{n=1}^{\infty}\frac{1}{n}
 \frac{{r_0}^n}{{r}^{n+1}}
 P_n(\cos\theta)=\nonumber\\
&-\frac{1}{r}\ln\left|\frac{1}{2}
 \left(1-\frac{r_0}{r}\cos\theta+
 \sqrt{1+\left(\frac{r_0}{r}\right)^2-2\frac{r_0}{r}
 \cos\theta}\right)\right| \ .\nonumber
\end{align}
\end{lemma}
{\bf Proof:} Substituting $t$ for $r_0$ in (\ref{A4}):
\begin{align}
&\int^{r_0}_{0}\left(\sum_{n=1}^{\infty}
\frac{t^{n-1}}{{r}^{n+1}} P_n(\cos\theta)\right)dt=\\
&\int^{r_0}_{0}\left(\frac{1}{t\sqrt{r^2+t^2-2 r t
\cos\theta}}-\frac{1}{t r}\right) dt \\
&\Leftrightarrow
\sum_{n=1}^{\infty}\frac{1}{n}
\frac{{r_0}^n}{{r}^{n+1}}
P_n(\cos\theta)=\\
&=-\frac{1}{r}\left(\ln\left|\frac{r}{t}-
\cos\theta+\frac{1}{t}\sqrt{r^2+t^2-2
r t \cos\theta}\right|+\ln t\right){\bigg\vert}_0^{r_0}\\
&=-\frac{1}{r}\ln\left|\frac{1}{2}\left(1-\frac{r_0}{r}\cos\theta+
\sqrt{1+\left(\frac{r_0}{r}\right)^2-
2\frac{r_0}{r}\cos\theta}\right)\right|.
\end{align}
Plugging $R^2$ instead of $r$ and $r r_0$ instead of $r$ in Lemma
(\ref{FA1}) yields:
\begin{align}\label{A9}
&\RE{\phi}^{0}=\notag\\
&\frac{e}{4\pi}\bigg(\frac{R}{\sqrt{R^4+{r_0}^2 r^2-2
r_0 r R^2\cos\theta}} -\frac{1}{R}+\frac{1}{R}\ln(2 R^2)-\notag\\
&\frac{1}{R}\ln\left|R^2-r_0 r\cos\theta+\sqrt{R^4+{r_0}^2
r^2-2 r_0 r R^2\cos\theta}\right|\bigg).
\end{align}
Figure  (\ref{rysunek:pole:monopola}) shows the directions of the
field $D - E = \Si{D}+\RE{D}^{0}
+\frac{e}{4\pi}\nabla\frac{1}{r}$. Observe that the field is
tangent to the boundary of $K(0,R)$.
\section{Renormalized energy for Neumann solutions}
\label{dodatek:energia}
To compute integral (\ref{calka:energii}):
\begin{align}
{\cal H}=&m- \frac{1}{2}\int_{{\Bbb
R^3}-V}\Si{D^2}dv+\frac{1}{2}\int_{V}\RE{D^2}
dv+\nonumber\\
&+\int_{V}\Si{D}\RE{D}dv \ ,
\end{align}
observe that:
\begin{gather}
 - \frac{1}{2}\int_{{\Bbb R^3}-V}\Si{D^2}dv=\frac{1}{2}
 \int_{{\partial
  \Bbb R^3}-\partial V}\Si{\phi}\Si{{D^{\bot}}}\,d\sigma
  \ ,\\
 \frac{1}{2}\int_{V}\RE{D^2}dv=-\frac{1}{2}\int_{\partial  V-
 {\partial  \Bbb R^3}} \RE{\phi}\RE{{D^{\bot}}}\,d\sigma \ .
\end{gather}
Integrals containing products of singular and regular fields are
understood in the sense of distributions (cf.~\cite{maurin}, p.
748). Denoting $k_{\epsilon}:=K(\vec{r}_0,\epsilon)$ we obtain:
\begin{gather}
\int_{V}\Si{D}\RE{D}dv=\lim_{\epsilon\rightarrow
0}\int_{V-k_{\epsilon}}\Si{D}\RE{D}dv=\notag\\
=-\lim_{\epsilon\rightarrow
0}\frac{1}{2}\int_{V-k_{\epsilon}}\nabla
\left(\Si{\phi}\RE{D}+\RE{\phi}\Si{D}\right)dv=\notag\\
=-\lim_{\epsilon\rightarrow 0}\frac{1}{2}\int_{\partial V-\partial
k_{\epsilon}}\left(\Si{\phi}\RE{{D^{\bot}}}+
\RE{\phi}\Si{{D^{\bot}}}\right)\,d\sigma \ .
\end{gather}
Hence,  for $V=K_R:=K(0,R)$ we have:
\begin{align}\label{H:epsilon:ogolny}
{\cal H}=&m-\frac{1}{2}\int_{\partial K_R}\phi{D^{\bot}}\,d\sigma+\nonumber\\
&+\lim_{\epsilon\rightarrow0}\frac{1}{2} \int_{\partial
k_{\epsilon}}\left(\RE{\phi}\Si{D^{\bot}}+
\Si{\phi}\RE{D^{\bot}}\right)\,d\sigma.
\end{align}
The formula is true for both the monopole and the dipole
singularity of $\Si{D}$. Here, we consider the monopole (Coulomb)
singularity. In this case the function $\Si{\phi}$ multiplied by
$\epsilon^2$ (coming from the surface measure $d\sigma$) vanishes
for $\epsilon \rightarrow 0$. Hence, we have:
\begin{equation}
{\cal H}=m-\frac{1}{2}\int_{\partial K_R}\phi{D^{\bot}}\,d\sigma+
\lim_{\epsilon\rightarrow0}\frac{1}{2} \int_{\partial
k_{\epsilon}}\RE{\phi}\Si{D^{\bot}}\,d\sigma \ .
\end{equation}
To compute the integral over $\partial  k_{\epsilon}$, we use
spherical coordinates $(\epsilon , \beta ,\varphi)$ centered at
$\vec{r}_0$. Parameters $r$ and $\cos\theta$ present in \RE{\phi}
may be expressed as follows:
\begin{align}\label{B7}
&r^2=r_0^2+{\epsilon}^2-2{\epsilon}r_0\cos\beta,\qquad
r\cos\theta=r_0-\epsilon\cos\beta \ ,\\ &
\lim_{\epsilon\rightarrow0} r^2=r_0^2,\qquad
\lim_{\epsilon\rightarrow0} r\cos\theta=r_0 \ .\label{B8}
\end{align}
Then:
\begin{align}
&\lim_{\epsilon\rightarrow0}\frac{1}{2} \int_{\partial
k_{\epsilon}}\RE{\phi}\Si{D^{\bot}}\,d\sigma=\nonumber\\
&\lim_{\epsilon\rightarrow0}\frac{e}{2}
\frac{2\pi}{4\pi}\int_{0}^{\pi}\frac{1}{\epsilon^2}\RE{\phi}
(r_0,\epsilon,\beta)\sin\beta \epsilon^2\,d\beta=\notag \\
&=\frac{e}{4}\RE{\phi}(r_0,r=r_0,\theta=0)\int_{0}^{\pi}\sin\beta
d\beta =\frac{1}{2}e\RE{\phi}{\big\vert}_{\vec{r}=\vec{r}_0} \ .
\end{align}
Consequently:
\begin{equation}
{\cal H}=m-\frac{1}{2}\int_{\partial
K_R}\phi{D^{\bot}}\,d\sigma+\label{hsing}
\frac{1}{2}e\RE{\phi}{\big\vert}_{\vec{r}=\vec{r}_0}.
\end{equation}
Knowing $\phi$ we can compute ${\cal H}$:
\begin{align}
&D^{\bot}{\vert}_{r=R}=\frac{e}{4\pi}\frac{1}{R^2}+E\cos\theta,\\
&{\phi}{\vert}_{r=R}=-E R\cos\theta+\notag\\
&+\frac{e}{4\pi}\bigg(
\frac{2}{\sqrt{R^2+{r_0}^2 -2 r_0 R \cos\theta}}
-\frac{1}{R}+\frac{1}{R}\ln(2 R)+\notag\\
&
-\frac{1}{R}\ln\left|R-r_0 \cos\theta+ \sqrt{R^2+{r_0}^2 -2 r_0 R
\cos\theta}\right|\bigg),\\
&e\RE{\phi}{\big\vert}_{\vec{r}=\vec{r}_0}=\nonumber\\
&=- e Er_0+\frac{e^2}{4\pi} \left(\frac{R}{R^2-r_0^2}
-\frac{1}{R}-\frac{1}{R}\ln\left|1-\frac{r_0^2}{R^2}
\right|\right).\label{h3}
\end{align}
Note that:
\begin{align}
&\int_{K_R}\frac{1}{\sqrt{R^2+{r_0}^2 -2 r_0 R \cos\theta}}
d\sigma=4\pi R,\\
& \int_{K_R}E\cos\theta d\sigma=0,\\
&\int_{K_R}\ln\left|R-r_0 \cos\theta+ \sqrt{R^2+{r_0}^2 -2 r_0
R\cos\theta}\right|d\sigma =\nonumber\\
&=4\pi R^2\ln 2 R,\notag
\end{align}
where we used  two integrals $2.736$ from \cite{rizik}. Then:
\begin{align}\label{h2}
\frac{e}{4\pi R^2}\int_{K_R}\phi d\sigma&=\frac{e^2}{4\pi R}.
\end{align}
Moreover:
\begin{align}
& E^2 R\int_{K_R}\cos^2\theta d\sigma=\frac{4}{3}\pi R^3 E^2,\\
& \int_{K_R}\frac{E\cos\theta}{\sqrt{R^2+{r_0}^2 -2 r_0 R
\cos\theta}} d\sigma=\frac{4}{3}\pi E r_0,\\
& \int_{K_R}\ln\left|R-r_0 \cos\theta+ \sqrt{R^2+{r_0}^2 -2 r_0
R\cos\theta}\right|\times\nonumber\\
&\times E\cos\theta d\sigma=-\frac{4}{3}\pi r_0 R ,
\end{align}
where we used  four integrals  $2.736$ from \cite{rizik}. Then:
\begin{align}\label{h1}
 E\int_{K_R}&\phi\cos\theta d\sigma=
 -\frac{4}{3}\pi R^3 E^2+\frac{eE}{4\pi}
 \left(\frac{8}{3}\pi r_0+\frac{4}{3}\pi r_0\right)=\nonumber\\
&= -\frac{4}{3}\pi R^3 E^2+ e E r_0.
\end{align}
The final result is the sum of (\ref{h3}), (\ref{h2}) and
(\ref{h1}) with coefficient $\frac{1}{2}$:
\begin{align}
{\cal   H}=&m+\frac{1}{2}\left(\frac{e^2}{4\pi}\left(\frac{R}{R^2-r_0^2}
-\frac{1}{R}\ln\left|1-\frac{r_0^2}{R^2}\right|-\frac{2}{R}\right)+\right.\nonumber\\
&+\left.\frac{4}{3}\pi R^3 E^2-2 e E r_0\right).
\end{align}
\section{Dirichlet solution and the corresponding energy}
\label{dodatek:pole:elektryczne:dirichlet}
To find a solution of the Poisson equation
$\Delta\phi=-e\delta(\vec{r}-\vec{r}_0)$ with boundary conditions
(\ref{Dir-co}),  where $\| \vec{r}_0 \| < R$ and $\vec{r}_0 \|
\vec{E}$, we denote:
$\phi=\Si{\phi}+\RE{\phi}^{0}-\vec{E}\vec{r}$, where $\Si{\phi} =
\frac{1}{4\pi}\frac{e}{|\vec{r}-\vec{r}_0|}$, $\Delta\RE{\phi}^{0}
=0$ and:
\begin{align}
\RE{\phi}^{0}{\big\vert}_{r=R}=-\frac{1}{4\pi}
\left(\frac{e}{|\vec{r}-\vec{r}_0|}\right)
{\bigg\vert}_{r=R}+\frac{e}{4\pi R} \ .
\end{align}
Again, we use {\em Ansatz} (\ref{A5}) as we did in Appendix
\ref{dodatek:pole:elektryczne}, page
\pageref{dodatek:pole:elektryczne}, and expand also boundary
conditions:
\begin{equation}
\RE{\phi}^{0}{\bigg\vert}_{r=R}=\frac{e}{4\pi}
\left(\frac{1}{r}-\frac{1}{|\vec{r}-\vec{r}_0|}\right)
{\bigg\vert}_{r=R} \ ,\label{Di6}
\end{equation}
in series of Legendre polynomials. After substitution (\ref{A4})
and (\ref{A5}) to (\ref{Di6}) we obtain:
\begin{equation}\label{Di7}
\RE{\phi}^{0}=-\frac{e R}{4\pi} \sum_{n=1}^{\infty}
\frac{{(r r_0)}^n}{{(R^2)}^{n+1}} P_n(\cos\theta).
\end{equation}
After rescaling (\ref{A4}) we get:
\begin{align}\label{C8}
\RE{\phi}^{0}=\frac{e}{4\pi}\left(\frac{1}{R}-\frac{R}
{\sqrt{R^4+{r_0}^2 r^2-2 r_0 r R^2\cos\theta}}\right).
\end{align}
Singular part of the electric field has the Coulomb singularity at
$\vec{r}_0$. Hence, formula (\ref{hsing}) is valid. However, we
have:
\begin{align}
D^{\bot}{\vert}_{r=R}&=\frac{e}{4\pi R}\frac{R^2-r_0^2}
{(R^2+{r_0}^2 -2 r_0  R\cos\theta)^{\frac{3}{2}}}+E\cos\theta,\\
{\phi}{\vert}_{r=R}&=-E R\cos\theta+\frac{e}{4\pi}\frac{1}{R},\\
e\RE{\phi}{\big\vert}_{\vec{r}=\vec{r}_0}&=- e E
r_0+\frac{e^2}{4\pi}
\left(\frac{1}{R}-\frac{R}{R^2-r_0^2}\right).\label{Dih3}
\end{align}
This implies:
\begin{align}
2\pi R^2 &\int_0^{\pi}\frac{e^2}{(4\pi
R)^2}\frac{(R^2-r_0^2)\sin\theta\,d\theta} {(R^2+{r_0}^2 -2 r_0
R\cos\theta)^{\frac{3}{2}}}=\frac{e^2}{4\pi R},\\ -2\pi R^2
&\int_0^{\pi} E R\cos\theta\,\frac{e}{4\pi
R}\frac{(R^2-r_0^2)\sin\theta\,d\theta} {(R^2+{r_0}^2 -2 r_0
R\cos\theta)^{\frac{3}{2}}}=\nonumber\\
&=-e E r_0,\\ -2\pi R^2 &\int_0^{\pi}
E^2 R \cos^2\theta\,\sin\theta\,d\theta=-\frac{4}{3}\pi R^3 E^2,\\
&\int_0^{\pi}\cos\theta\,\sin\theta\,d\theta=0.
\end{align}
Consequently, we obtain:
\begin{gather}
 {\cal H}=m+\frac{1}{2}\left(\frac{4}{3}\pi R^3 E^2
 -\frac{e^2}{4\pi}\frac{R}{R^2-r_0^2}\right) \ ,
\end{gather}
or, in standardized variables (\ref{standard}),
\begin{equation}
{\cal H_{D}'}=\frac{1}{3} q^2-\frac{1}{1-x^2}.
\end{equation}
\section{Dipole particle in a spherical box}
\label{dodatek:pole:dipola}
We must solve equation $\Delta\phi^{dip}  =
\vec{p}\cdot\nabla\,(\delta(\vec{r}-\vec{r}_0))$ with boundary
conditions $\vec{D}^{dip}\cdot\vec{n}{\big\vert}_{r=R}=0$.
Denoting $\phi=\Si{\phi}^{dip}+\RE{\phi}^{dip}$, where
\begin{equation}\label{dip-sing}
  \Si{\phi}^{dip} =
   \frac{1}{4\pi}\frac{\vec{p}\cdot(\vec{r}-
   \vec{r}_0)}{|\vec{r}-\vec{r}_0|^3} \ ,
\end{equation}
we get Laplace equation $\Delta\RE{\phi}^{dip} =0$ with boundary
condition:
\begin{equation}\label{P3}
  \RE{\vec{D}}^{dip}\cdot\vec{n}{\big\vert}_{r=R}
  =\vec{n}\cdot\frac{1}{4\pi}\nabla
  \left(\frac{\vec{p}\cdot(\vec{r}-\vec{r}_0)}
  {|\vec{r}-\vec{r}_0|^3}\right)
  {\bigg\vert}_{r=R} \ .
\end{equation}
For any pair of vectors $\vec{r}_0$ i  $\vec{p}$ we choose
coordinates in which   $\vec{r}_0$ is parallel to the $z$-axis
${\bf e}_z$ and polarization vector assumes the form
$\vec{p}=p{\boldsymbol e_z}+p_x{\boldsymbol e_x}$. The final
solution will be the sum of two harmonic functions fulfilling
boundary condition (\ref{P3}), calculated \emph{separately} for
$p{\boldsymbol e_z}$ and $p_x{\boldsymbol e_x}$.

Observe that, for $\RE{\phi^{mon}}(\vec{r}_0 , \vec{r})$ being a
solution of Laplace equation, also the function $\frac{\vec{p}}e
\cdot \nabla_{\vec{r}_0} \RE{\phi^{mon}}$ is harmonic. Moreover,
if \RE{\phi^{mon}} fulfills conditions ((\ref{A6}) condition from
page \pageref{A6}):
\begin{equation}
 \frac{\partial}{\partial
 r}\RE{\phi^{mon}}(\vec{r} ,\vec{r}_0){\bigg\vert}_{r=R}=
 \frac{e}{4\pi}\frac{\partial}{\partial r}
 \left(\frac{1}{r}-\frac{1}{|\vec{r}-\vec{r}_0|}\right)
 {\bigg\vert}_{r=R} \ ,
\end{equation}
then, after differentiation with respect to $\vec{r}_0$ we obtain:
\begin{align}
-\frac{\partial}{\partial r}&\left(\frac{\vec{p}}e \cdot
\nabla_{\vec{r}_0}\RE{\phi^{mon}}\right)
{\bigg\vert}_{r=R}=&\nonumber\\
&=\frac{1}{4\pi}\frac{\partial}{\partial r}\left(
\vec{p} \cdot \nabla_{\vec{r}_0}
\frac{1}{|\vec{r}-\vec{r}_0|}
\right) {\bigg\vert}_{r=R} \ .
\end{align}
Hence, the function $\frac{\vec{p}}e \cdot \nabla_{\vec{r}_0}
\RE{\phi^{mon}}$ satisfies boundary conditions (\ref{P3}). We
conclude that:
\begin{align}\label{P5}
\RE{\phi}^{dip}=\frac{1}{4\pi e}\left(\vec{p}\cdot
\nabla_{\vec{r}_0}\right)\RE{\phi}^{mon} \ ,
\end{align}
(cf.~\cite{Panofsky}, p.14). Applying (\ref{P5}) for
$\vec{p}=p{\boldsymbol e_z}+p_x{\boldsymbol e_x}$ allows us to
solve the problem separately for $p$ parallel and orthogonal to
$\vec{r}_0$.
\subsection{Solution for $\vec{p}\parallel \vec{r}_0$}
\begin{figure}
\centering
\epsfig{file=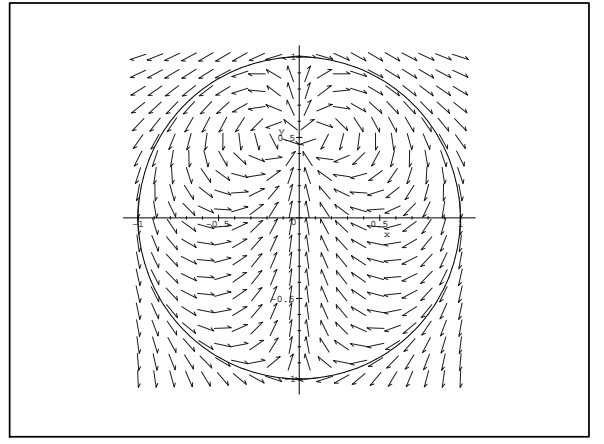,height=8.5cm, angle=-90}
\caption{Directions of the field $\Si{D}^{dip}+\RE{D}^{dip}$,
 $R=1$, $r_0=0.5$, $p=1$}
\label{rysunek:pole:dipola}
\end{figure}

To obtain the parallel part  we differentiate monopole solution
((\ref{A9}), Appendix \ref{dodatek:pole:elektryczne}) along the
${\bf e}_z$-axis:
\begin{align}
&\RE{\phi}^{dip}=\frac{p}{e}\,\frac{\partial}{\partial
r_0}\RE{\phi^{mon}}=\nonumber\\
&\frac{p}{4\pi}\frac{\partial}{\partial r_0}
\bigg(\frac{R}{\sqrt{R^4+{r_0}^2 r^2-2 r_0 r R^2\cos\theta}}
-\frac{1}{R}+\frac{1}{R}\ln(2 R^2)\notag\\ &
-\frac{1}{R}\ln\left|R^2-r_0 r\cos\theta+\sqrt{R^4+{r_0}^2 r^2-2
r_0 r R^2\cos\theta}\right|\bigg)\notag\\
&=\frac{p}{4\pi}\bigg(-\frac{R(r_0 r^2-r R^2\cos\theta)}
{(R^4+{r_0}^2 r^2-2 r_0 r R^2\cos\theta)^{\frac{3}{2}}}+\nonumber\\
&-\frac{1}{R\sqrt{R^4+{r_0}^2 r^2-2
r_0 r R^2\cos\theta}}\times\notag
\\
&\frac{-r\cos\theta\left(\sqrt{R^4+{r_0}^2 r^2-2 r_0 r
R^2\cos\theta}+R^2\right)+r^2 r_0}{R^2-r
r_0\cos\theta+\sqrt{R^4+{r_0}^2
r^2-2 r_0 r R^2 \cos\theta}}\bigg).
\end{align}
But:
\begin{align}
&\left(R^2-r r_0\cos\theta+\sqrt{R^4+{r_0}^2 r^2-2 r_0 r R^2
\cos\theta}\right)\times\label{ppomocnicze}\\
&\times\left(R^2-r r_0\cos\theta-\sqrt{R^4+{r_0}^2 r^2-2 r_0 r R^2
\cos\theta}\right)=\nonumber\\
&=-(r_0 r)^2\sin^2\theta,\\
&\left(-r\cos\theta\left(\sqrt{R^4+{r_0}^2 r^2-2
r_0 r R^2\cos\theta}+R^2\right)+r^2 r_0\right)\nonumber\\
&\times\left(R^2-r r_0\cos\theta-\sqrt{R^4+{r_0}^2 r^2-2 r_0 r R^2
\cos\theta}\right)=\nonumber\\
&=-r_0 r^2\sin^2\theta \sqrt{R^4+{r_0}^2 r^2-2 r_0 r R^2
\cos\theta}+\nonumber\\
& r\cos\theta(-r_0r R^2\cos\theta)+R^2r^2 r_0=\nonumber\\
&=r^2 r_0\sin^2\theta\left(R^2-\sqrt{R^4+{r_0}^2
r^2-2 r_0 r R^2 \cos\theta}\right).
\end{align}
So:
\begin{align}
&\RE{\phi}^{dip}=\frac{p}{4\pi}\bigg(-\frac{R(r_0 r^2-r
R^2\cos\theta)}
{(R^4+{r_0}^2 r^2-2 r_0 r R^2\cos\theta)^{\frac{3}{2}}}+\notag\\
&+\frac{r^2 r_0\sin^2\theta\,(R^2-\sqrt{R^4+{r_0}^2 r^2-2 r_0
r R^2 \cos\theta})}{R(r r_0)^2\sin^2\theta\,\sqrt{R^4+{r_0}^2
r^2-2 r_0 r R^2 \cos\theta}}\bigg)=\notag\\
&=\frac{1}{4\pi}\left(\frac{p\,R^3\left(R^2-r
r_0\cos\theta\right)}{r_0 \left(R^4+(r_0 r)^2-2r r_0
R^2\cos\theta\right)^{\frac{3}{2}}} -\frac{p}{r_0 R}\right).
\end{align}
Figure \ref{rysunek:pole:dipola} shows the directions of the field
$D^{dip}$. Observe that the field is tangent to the boundary of
$K(0,R)$.
\subsection{Solutions for $\vec{p}\bot\vec{r_0}$}
For $\vec{p}=p_x{\boldsymbol e_x}$ we get:
\begin{equation}
\RE{\phi}^{dip}= p_x\frac{{\boldsymbol e_x}}e \cdot
\nabla_{\vec{r}_0} \RE{\phi}^{mon}=
\frac{p_x}{e}\frac{\partial}{\partial x_0}\RE{\phi}^{mon}.
\end{equation}
The easiest way to calculate this derivative is to use spherical
coordinates $\vec{r}_0=(r_0,\theta_0,\varphi_0)$. Then:
\begin{align}
x_0&=r_0\sin\theta_0\sin\varphi_0\& ,\quad
y_0=r_0\sin\theta_0\cos\varphi_0\ ,\nonumber\\
 z_0&=r_0\cos\theta_0\ ,
\end{align}
and:
\begin{align}
\frac{\partial}{\partial
x_0}&=\sin\theta_0\cos\varphi_0\frac{\partial}{\partial r_0}+
\cos\theta_0\cos\varphi_0\frac{1}{r_0}\frac{\partial}{\partial
\theta_0}+\nonumber\\
&-\frac{\sin\varphi_0}{r_0\sin\theta_0}\frac{\partial}{\partial
\varphi_0} \ . \label{D13}
\end{align}
But for $\vec{r}_0 \| {\bf e}_z$ this procedure is singular
because $\sin\theta_0 =0$. To overcome this difficulty we first
calculate the result for $\vec{r}_0 \nparallel {\bf e}_z$ and then
pass to the limit $\theta_0 \rightarrow 0$ and $\varphi_0
\rightarrow 0$. For this purpose we must be able to differentiate
the function $\cos \gamma$, where $\gamma$ is the angle between
$\vec{r}$ and $\vec{r}_0$, i.e.:
\begin{equation}
\vec{r}\cdot\vec{r}_0=r     r_0     \cos\gamma \ ,
\end{equation}
or, equivalently:
\begin{equation}
\cos\gamma=\cos\theta\cos\theta_0+
\sin\theta\sin\theta_0\cos(\varphi-\varphi_0) \ .
\end{equation}
Hence, (\ref{D13}) gives us:
\begin{align}
&\frac{\partial}{\partial x_0} \cos\gamma=\nonumber\\
&\bigg(
\frac{1}{r_0}\cos\theta\cos\varphi_0\left(-\cos\theta\sin\theta_0+
\sin\theta\cos\theta_0\cos(\varphi-\varphi_0)\right)\notag\\
&-\frac{1}{r_0}\sin\varphi_0\sin\theta\sin(\varphi-\varphi_0)\bigg)
\overset{\theta_0\rightarrow 0}
{\underset{\varphi_0\rightarrow 0}{\longrightarrow}}
 \frac{1}{r_0}\sin\theta\cos\varphi.
\end{align}
This method allows us to calculate effectively the derivative of
the monopole field from Appendix \ref{dodatek:pole:elektryczne}
(p. \pageref{dodatek:pole:elektryczne}) along $x_0$. The final
result reads:
\begin{align}
&\RE{\phi}^{dip}=\frac{p_x}{4\pi}\Bigg(\frac{r R^3\sin\theta\cos\varphi}
{(R^4+{r_0}^2 r^2-2 r_0 r R^2\cos\theta)^{\frac{3}{2}}}+\notag\\
&-\frac{\cos\varphi(R^2\cos\theta-r_0 r)}{R r_0\sin\theta
\sqrt{R^4+{r_0}^2 r^2-2 r_0  r
R^2\cos\theta}}+\frac{\cos\theta\cos\varphi}
{R r_0\sin\theta}\Bigg).
\end{align}
We stress that the above function is regular at $\theta = 0$ due
to cancellations between the second and the third term.

\section{Renormalized energy of a dipole particle}
\label{dodatek:energia:dipolowa}
To calculate ${\cal H}^{dip}$ we use results of Appendix
\ref{dodatek:energia}. It turns out that in formula
(\ref{H:epsilon:ogolny}), only the following non-vanishing terms
were not taken into account in ${\cal H}^{mon}$:
\begin{align}
&{\cal H}^{dip}=-\frac{1}{2}\int_{\partial
K_R}\phi^{dip}{D^{\bot}}\,d\sigma+\nonumber\\
&+\lim_{\epsilon\rightarrow0}\frac{1}{2} \int_{\partial
k_{\epsilon}}\left(\RE{\phi^{dip}}\Si{D^{\bot\, mon}} +
\Si{\phi^{dip}}\RE{D^{\bot}} \right)\,d\sigma  \ ,
\end{align}
where:
\begin{align}
&\RE{\phi^{dip}}=\frac{1}{4\pi}\left(\frac{p\,R^3\left(R^2-r
r_0\cos\theta\right)}{r_0 \left(R^4+(r_0 r)^2-2r r_0
R^2\cos\theta\right)^{\frac{3}{2}}} -\frac{p}{r_0
R}\right),\label{D1}\\
&\Si{\phi^{dip}}=\frac{1}{4\pi}\frac{p(r\cos\theta-r_0)}
{(r^2+r_0^2-2r r_0\cos\theta)^{\frac{3}{2}}},\\
&\RE{\phi^{mon}}=\frac{e}{4\pi}\bigg(\frac{R}{\sqrt{R^4+{r_0}^2
r^2-2 r_0 r R^2\cos\theta}} -\frac{1}{R}+\notag\\
 &-\frac{1}{R}\ln\left|R^2-r_0 r\cos\theta+
\sqrt{R^4+{r_0}^2 r^2-2 r_0 r R^2\cos\theta}\right|\bigg)\notag\\
&-Er\cos\theta+\frac{1}{R}\ln(2
R^2),\\
&\phi^{dip}=\RE{\phi^{dip}}+\Si{\phi^{dip}} \ .
\end{align}
Moreover, we have:
\begin{eqnarray}\label{D:prost}
  D^{\bot}{\big|}_{\partial K_R}&
   =& \frac{1}{4\pi}\frac{e}{R^2}+E\cos\theta \ , \\
\label{D2}
  \RE{D^{\bot}}{\big|}_{\partial  k_\epsilon}
  &=&-\frac{\partial}{\partial\epsilon}
\left(\RE{\phi^{mon}}+\RE{\phi^{dip}}\right) \ .
\end{eqnarray}
To compute the integral over  $\partial K_R$ we note that:
\begin{align}
&\phi^{dip}{\bigg\vert}_{r=R}=\nonumber\\
&\RE{\phi^{dip}}+\Si{\phi^{dip}}=
\frac{p}{4\pi r_0}\left(\frac{R^2-r_0^2}{(R^2+r_0^2-2R
r_0\cos\theta)^{\frac{3}{2}}}-\frac{1}{R}\right),
\end{align}
whereas $D^{\bot}$ is expressed by (\ref{D:prost}). Moreover:
\begin{equation}
\frac{1}{4\pi}\frac{e}{R^2}
2\pi\int_{0}^{\pi}\phi^{dip}\sin\theta\,d\theta=0.
\end{equation}
So:
\begin{align}
-\frac{1}{2}
2\pi\,E\int_{0}^{\pi}\phi^{dip}
\cos\theta\sin\theta\,d\theta=-\frac{1}{2}
p\,E.\notag
\end{align}
To find the limit:
\begin{equation}
\lim_{\epsilon\rightarrow0}\frac{1}{2}
\int_{\partial k_{\epsilon}}\left(\Si{\phi^{dip}}\RE{D^{\bot}}+
\Si{D^{mon}}\RE{\phi^{dip}}\right)\,d\sigma,
\end{equation}
we analyze behaviour of fields (\ref{D1}) - (\ref{D2}) for
$\epsilon\rightarrow 0$. All these terms have at most the
$\epsilon^{-2}$-singularity. Therefore, they are continuous and
bounded when multiplied by $\epsilon^2$. Thus, we can interchange
the limit and the integration operations.

We follow our procedure described in Appendix
\ref{dodatek:energia}, page \pageref{dodatek:energia}. Using
(\ref{B7}) and (\ref{B8}) we obtain in terms of the standardized
variable $x=\frac{r_0}{R}$:
\begin{align}
&\lim_{\epsilon\rightarrow0}\left(\epsilon^2\,\Si{\phi^{dip}}\right)=
-\frac{p}{4\pi}\cos\beta,\\
&\lim_{\epsilon\rightarrow0}\left(-\frac{\partial}{\partial\epsilon}
\RE{\phi^{mon}}\right)=\nonumber\\
&\frac{e}{4\pi}\left(\frac{r_0
R}{(R^2-r_0^2)^2}+\frac{r_0}{(R^2-r_0^2)^2 R}\right)\cos\beta
-E\cos\beta=\notag\\
&\left(\frac{e}{4\pi} \frac{1}{R^2}\frac{x(2-x^2)}{(1-x^2)^2}
-E\right)\cos\beta=
-\cos\beta\,\RE{D^{mon}}{\bigg\vert}_{\vec{r}=\vec{r_0}},\\
&\lim_{\epsilon\rightarrow0}\left(-\frac{\partial}{\partial\epsilon}
\RE{\phi^{dip}}\right)=
\frac{1}{4\pi}\frac{2 p
R^3}{(R^2-r_0^2)^3}\cos\beta=\notag\\
&=\frac{1}{4\pi}\frac{2
p}{R^3(1-x^2)^3}\cos\beta=
-\cos\beta\,\RE{D^{dip}}{\bigg\vert}_{\vec{r}=\vec{r_0}},\\
&\lim_{\epsilon\rightarrow0}\left(\epsilon^2\,\Si{D^{mon}}\right)
=\frac{e}{4\pi},\\
&\lim_{\epsilon\rightarrow0}\left(\RE{\phi^{dip}}\right)=
\frac{1}{4\pi}\frac{p
r_0(2R^2-r_0^2)}{R(R^2-r_0^2)^2}=\nonumber\\
&\frac{p}{4\pi}
\frac{x(2-x^2)}{R^2(1-x^2)^2}=
-\frac{p}{e}\RE{D^{mon}}{\bigg\vert}_{\vec{r}=\vec{r_0}},\\
&\int_0^{\pi}\cos^2\beta\sin\beta\,d\beta=\frac{2}{3}.
\end{align}
Then:
\begin{align}
&\frac{1}{2} \int_{\partial
k_{\epsilon}}\left(\Si{\phi^{dip}}\RE{D^{\bot}}+
\Si{D^{mon}}\RE{\phi^{dip}}\right)\,d\sigma=\notag\\
&=\frac{1}{2}\left(\frac{4\pi}{3}\frac{p}{4\pi}\left(
\RE{D^{mon}}{\bigg\vert}_{\vec{r}=\vec{r_0}}+
\RE{D^{dip}}{\bigg\vert}_{\vec{r}=\vec{r_0}}\right)\right.+\nonumber\\
&\left.-\frac{e}{4\pi}
4\pi\frac{p}{e}\RE{D^{mon}}{\bigg\vert}_{\vec{r}
=\vec{r_0}}\right)=\notag\\
&=\frac{1}{2}
\left(\frac{pe}{4\pi}\frac{1}{R^2}\frac{2}{3}\frac{x(2-x^2)}{(1-x^2)^2}
-\frac{1}{4\pi}\frac{1}{3}\frac{2
p^2}{R^3(1-x^2)^3}+\frac{1}{3}p\,E\right) .
\end{align}
Using $q=\frac{4\pi R^2}{e} E$ and (\ref{H-standard}) we obtain:
\begin{widetext}
\begin{align}
&{\cal H}^{\prime\ dip}:=\frac{8\pi R}{e^2}{\cal H}^{dip}
=\frac{2}{3} \left(\frac{p}{ e R}\frac{x(2-x^2)}{(1-x^2)^2}
-\frac{p^2}{e^2 R^2}\frac{1}{(1-x^2)^3}-\frac{p}{e R}\,q\right).
\label{energia:dodatkowa}
\end{align}\
\end{widetext}

\end{document}